\documentstyle[11pt,amsfonts,amssymb,amsmath,amscd]{article}
\begin{document}
\title{A Note on the Symmetries and Renormalisability of (Quantum) Gravity}
\author{Ioannis P. ZOIS
\\
School of Mathematics,
\\
Cardiff University, PO Box 926,
\\
 Cardiff CF24 4YH,
\\
UK.
\\
e-mail: zoisip@cf.ac.uk}

\newtheorem{thm}{Theorem}
\newtheorem{defn}{Definition}
\newtheorem{prop}{Proposition}
\newtheorem{lem}{Lemma}
\newtheorem{cor}{Corollary}
\newtheorem{rem}{Remark}
\newtheorem{ex}{Example}

\newcommand{\Rat}{\mathbb Q}
\newcommand{\Real}{\mathbb R}
\newcommand{\RR}{\Real}
\newcommand{\Rh}{\hat{\Real}}
\newcommand{\Nat}{\mathbb N}
\newcommand{\Complex}{\mathbb C}
\newcommand{\HH}{\mathbb H_3}
\newcommand{\CC}{\Complex}
\newcommand{\Z}{\mathbb Z}

\newcommand{\Ea}{{\mathcal E}}
\newcommand{\Ta}{{\mathcal A}}
\newcommand{\Aa}{{\Ta_\infty}}
\newcommand{\Eb}{{C^*(D_1,D_2,X_2,\Omega)}}
\newcommand{\Tb}{{C^*(D_1,D_2,\hat\Omega)}}
\newcommand{\Ab}{{C^*(D_1,D_2,\Omega)}}
\newcommand{\aco}{\idl}
\newcommand{\aca}{\beta^{\|}}
\newcommand{\acb}{\beta^\perp}
\newcommand{\acc}{{\hat\tau}}

\newcommand{\Uu}{{\cal U}}
\newcommand{\Dd}{{\mathcal D}}
\newcommand{\Oo}{{\mathcal O}}
\renewcommand{\H}{{\mathcal H}}
\newcommand{\NN}{{\bf N}}
\newcommand{\ZZ}{{\bf Z}}
\newcommand{\Pp}{{\mathcal P}}
\newcommand{\Zz}{{\mathcal Z}}
\newcommand{\PP}{{\bf P}}
\newcommand{\LL}{\Lambda}
\newcommand{\LLL}{\Lambda\cup \infty}
\newcommand{\EE}{{\bf E}}
\newcommand{\Bb}{{\mathcal B}}
\newcommand{\Ww}{{\mathcal W}}
\newcommand{\Ss}{{\mathcal S}}

\newcommand{\Tr}{\mbox{\rm Tr}}  
\newcommand{\TV}{{\mathcal T}}
\newcommand{\TVh}{\hat{{\mathcal T}}}
\newcommand{\tr}{\mbox{tr}}

\newcommand{\Rr}{{\mathcal R}}
\newcommand{\Nn}{{\mathcal N}}
\newcommand{\Cc}{{\mathcal C}}
\newcommand{\Jj}{{\mathcal J}}
\newcommand{\Ff}{{\mathcal F}}
\newcommand{\Ll}{{\mathcal L}}
\newcommand{\id}{{\mbox{\rm id}}}
\newcommand{\idl}{{\mbox{\rm\tiny id}}}
\newcommand{\eva}{{\mbox{\rm ev}}}
\newcommand{\eval}{{\mbox{\rm\tiny ev}}}

\def\essinf{\mathop{\rm ess\,inf}}
\def\esssup{\mathop{\rm ess\,sup}}

\newcommand{\bew}{{\bf Proof:}}
\newcommand{\eb}{\hfill $\Box$}

\newcommand{\x}{{\vec x}}
\newcommand{\y}{{\vec y}}
\newcommand{\n}{{\vec n}}
\renewcommand{\a}{{\vec a}}
\newcommand{\hull}{\Sigma}
\newcommand{\om}{\omega}
\newcommand{\oh}{{\hat{\om}}}

\newcommand{\Af}{{\Aa}_0}
\newcommand{\Tf}{{\Ta}_0}
\newcommand{\Ef}{{\Ea}_0}
\renewcommand{\H}{{\mathcal H}}

\newcommand{\bs}{\bigskip}
\newcommand{\ms}{\medskip}

\newcommand{\erz}[1]{\langle{#1}\rangle}
\newcommand{\pair}[2]{\erz{#1,#2}}
\newcommand{\diag}[2]{\mbox{\rm diag}(#1,#2)}
\newcommand{\tOmega}{{\Omega^s}}
\newcommand{\td}{{d^s}}
\newcommand{\tint}{{\int^s}}
\newcommand{\ttau}{{\tilde\tau}}
\newcommand{\talpha}{{\tilde\alpha}}
\newcommand{\tS}{{\tilde S}}
\newcommand{\ual}{{\underline{\alpha}}}
\newcommand{\hotimes}{{\hat\otimes}}
\newcommand{\K}{{\mathcal K}}
\newcommand{\Ypsilon}{{\Theta}}
\newcommand{\CA}{$C^*$-algebra}
\newcommand{\CF}{$C^*$-field}
\newcommand{\G}{\mathcal G}
\newcommand{\im}{\mbox{\rm im\,}}
\newcommand{\rk}{\mbox{\rm rk\,}}
\newcommand{\cp}{{\rtimes}}
\newcommand{\del}{{\bf \delta}}

\renewcommand{\a}{{\vec a}}
\newcommand{\at}{{\bf \tilde a}}
\newcommand{\cy}{\Phi}
\newcommand{\co}{\cy_\a}
\newcommand{\tco}{\cy_{\at}}

\newcommand{\ttimes}{{\tilde \rtimes}}
\renewcommand{\ss}{{\mathcal R}}
\renewcommand{\Cc}{{\mathcal C}}
\newcommand{\Ri}{(\RR\cup\infty)}
\newcommand{\Rd}{\hat\RR}
\newcommand{\dom}{\mbox{\rm dom}}
\newcommand{\fin}{\mbox{\rm\tiny fin}}

\newcommand{\enn}{\mu}
\newcommand{\ch}{\mbox{\rm ch}}
\newcommand{\supp}{\mbox{\rm supp}}
\newcommand{\chd}{\ch}
\newcommand{\Oh}{\hat\hull}

\newcommand{\hsp}{\RR^{d-1}\times\RR^{\leq 0}}
\newcommand{\IDS}{IDS}
\newcommand{\sv}[2]{\left(\begin{array}{c} #1 \\ #2 \end{array}\right)}

\bibliographystyle{amsalpha}

\maketitle

\begin{abstract}

We make some remarks on the group of symmetries in gravity; we believe that K-theory and noncommutative geometry 
inescepably have to play an important role. Furthermore we make some comments and questions on the recent work of 
Connes and Kreimer on renormalisation, the Riemann-Hilbert correspondence and their relevance to quantum gravity.\\



\end{abstract}

\section{Introduction and Motivation}

The main reason for the prominent role of Yang-Mills theories in physics is the fact that such \textsl{quantum}
 theories make sense, namely one can extract \textsl{finite} answers for physical quantities
 through a process known as \emph{renormalisation}. The proof of this celebrating fact was given some 30 years ago by 
G. 't Hooft who has also introduced the method of \textsl{dimensional regularisation} (see for instance \cite{iz}). 
Regularisation is the first step in the 2-step process of renormalisation where one wants to "parametrise the infinities" 
which appear in quantum field theories; explicitly one wants to calculate 
divergent integrals of the form
$$\int_{0}^{\infty}d^{4}kF(k)$$
 (where $k$ is essentially the momentum). We know that the 3 out of the 4 known interactions in nature are 
Yang-Mills theories. Thus one can indeed have a meaningful 
quantum theory for electroweak and strong interactions.\\

Gravity however, the 4th known interaction in nature, is a different story: although it can be thought of as a gauge theory, 
it is not of Yang-Mills type since it has a different action and a different gauge group of symmetries.
From the early days of the development of quantum field theory (due to Dirac, Schwinger, Dyson, Feynman etc), people knew 
that gravity suffers from \emph{(incurable perhaps)} divergencies and infinities and all known methods of regularisation 
which worked in other theories, such as the Pauli-Villars method, the momentum cutoff method or dimensional regularisation, 
they all brake down in this case. Of course since the days 
of A. Einstein, a quantum theory of gravity is every self-respectful physicist's dream. It is perhaps surprising the fact 
that although gravity is the weakest of all interactions and one might expect perturbative methods to work quite well
for it, it is the interaction for which all known renarmalisation schemes fail. Being optimistic, we shall not call
 gravity a nonrenormalisable theory, we shall say that it is only \textsl{"superficially nonrenormalisable"} 
in dimension 4 since the upper 
critical dimension of Newton's constant $G$ is 2 as follows from the relevant Callan-Symanzik equation 
(see section 4 below). In such cases, which are not at all promising,
what can \textsl{perhaps} save the day is an elaborate symmetry argument, certain fixed points of the renormalisation 
group flow or nonperturbative effects. That's the main motivation for this 
piece of work.\\ 

{\bf Aside:} We take the point of view that quantum gravity-which is currently an elusive
theory-\emph{should exist;} the argument in favour of its existence goes as follows (the original argument we think was
due to P.A.M. Dirac): let us consider Einstein's classical field equations 
which describe gravity (we assume no cosmological constant and we use physical units, i.e. we set the speed of light 
and Planck's constant equal to one):
$$G_{\mu\nu}=8\pi GT_{\mu\nu}$$
In the above equation, $G$ denotes Newton's constant, $T_{\mu\nu}$ denotes the energy-momentum tensor and 
$G_{\mu\nu}$ denotes the Einstein 
tensor which is equal, by definition, to $G_{\mu\nu}:=R_{\mu\nu}-\frac{1}{2}Rg_{\mu\nu}$, where $g_{\mu\nu}$ is the 
Riemannian metric, $R_{\mu\nu}$ 
is the Ricci curvature tensor and $R$ is the scalar curvature. One can see clearly that the RHS of the above
equation,namely the energy-momentum tensor, contains 
mass and energy coming from the other two interactions in nature (electroweak and strong); mass for instance, consists 
primarily of
quarks and leptons (these are both fermions); one also has the massive carriers of the electroweak force, the W and Z 
bosons; 
they all aquire mass through the Higgs mechanism (the Higgs boson--the carriers of the strong force and electromagnetism,
namely the gluons and the photons, are massless). We know that these interactions (strong and electroweak) are quantized
 and hence
the RHS of the equation contains \emph{quantized quantities}. So for \emph{consistency} of the equations, the 
LHS, which encodes geometry, \textsl{should also be quantized}.\\

[Comment: one may argue that the LHS may remain 
classical while the RHS may involve the \textsl{average value} of an operator; however such a theory will not be 
essentially different from classical general relativity and probably not qualified to be called quantum gravity, what we 
have in mind is Ehrenfert Theorem from Quantum Mechanics. We think of the above 
field equations as describing, in the quantum level, an actual equality between operators].\\

\section{Gravity and Yang-Mills Theories}

Let us elaborate more on the two differences between gravity and Yang-Mills theories (like the strong and the 
electroweak forces) at the \textsl{classical level}.
The first difference is the action: in Yang-Mills theories we start with a 4-dim (pseudo) 
Riemannian manifold $M$
 represetning spacetime along with a structure Lie group $G$ (say $G$ is some $SU(N)$; to be phenomenologically correct, 
$G=SU(3)$ for the strong force and $G=U(2)$ for the electroweak force) representing internal symmetries; we thus construct
 a principal $G$-bundle over $M$ whose total space $P$ gives the internal space of the theory; we pick a connection $A$ on 
the bundle $P$ (which represents the gauge potential) with curvature (field strength) 
$F:=d_{A}A:=dA-\frac{1}{2}[A,A]$ where 
$d_{A}$ denotes the exterior covariant derivative
 w.r.t. the
 conection 1-form $A$. Then the (pure) Yang-Mills action reads (ignoring constants)
$$I=\int_{M}F\wedge *F$$
where "$*$" denotes the Hodge dual which is defined using the Riemannian metric. The group of (internal) gauge 
transformations is the infinite dim Lie group of bundle 
automorphisms, denoted $B$, covering the identity map on the base manifold--sometimes these are called 
\emph{strong} bundle automorphisms, (or equivalently $B=Maps(M\rightarrow G)$) (see \cite{atiyah2}).
 The Euler-Lagrange equations read
$$d_{A}*F="source"$$
The above equations state the deep geometric fact that \textsl{"the curvature of the internal space is caused by the
 existence of the relevant charges"}. Similarly the corresponding monopoles are singular points where the Bianchi
 identity fails.\\

Gravity is different: the (Einstein-Hilbert) action reads
$$I=\int_{M} R$$
where $R$ denotes the scalar curvature of the Levi-Civita connection defined via the metric. The corresponding 
Euler-Lagrange equations are Einstein's equations which (ignoring constants) equate the Einstein tensor with the
 energy-momentum tensor. These equations are different (but in similar spirit) from the Yang-Mills equations: the
 internal space is spacetime itself (or its tangent bundle
 to be more precise) and the relevant charge for the gravitational interaction is mass. Einstein's equations
 then tell us qualitatively that \textsl{"it is not only mass (ie the relevant charges) which curve the internal
 space but there is additional curvature coming from the intrinsic geometry of the spacetime manifold itself"}.
 As about the group of (spacetime) gauge transformations, this is the infinite dim Lie group of local diffeomorphisms
 $Diff(M)$ of $M$.\\

Thus the total group of symmetries, denoted $T$, is the semi-direct product $T=B\times Diff(M)$. The situation is 
summarised by the following exact sequence of groups:
$$1\rightarrow B\rightarrow T\rightarrow Diff(M)\rightarrow 1$$
Clearly in order to unify strong and electroweak forces we should take $B$ to 
be the group of strong bundle maps with strucure group $G=U(2)\times SU(3)$.\\

If one wishes to build a unifying theory of all interactions, there are two obvious ways to
 proceed: One can either try to see if there is a "space" $\tilde{M}$ such that $T=Diff(\tilde{M})$.
This means that at least as far as symmetries are concerned, we would like to make the would-be unified theory 
actually "look like" a "gravity" theory on a new spacetime manifold $\tilde{M}$. This approach was adopted by Connes 
et all (see \cite{connes}) and it is useful if one wants to use the quantum theory in order to reveal the deep underlying
 "quantum" geometry of spacetime; 
 by following this approach one ends up with the Connes-Lott model and its variations (the double-sheeted spacetime),
 namely the new spacetime $\tilde{M}$ is a \emph{noncommutative} space (ie a space whose algebra of coordinate
 functions is noncommutative) where the metric is given by the inverse of the Dirac operator (Dirac propagator)
 or the Schwinger-Dyson propagator used more recently in \cite{connes1}. In noncommutative geometry one replaces the group 
$Diff(M)$ by the automorphism group $Aut(A)$ of a noncommutative algebra $A$. This follows from Gelfand's theorem and 
from the exact sequence of groups
$$1\rightarrow Int(A)\rightarrow Aut(A)\rightarrow Out(A)$$
where $Int(A)$ and $Out(A)$ denote the groups of internal and external automorphisms respectively of the algebra $A$.\\

The second way is to try to see if there is a suitable "extended" (Lie perhaps) group $\tilde{G}$ which we use in order 
to construct a principal $\tilde{G}$-bundle with total space $\tilde{P}$ over ordinary spacetime $M$ such that $T$
 equals the Lie group of strong $\tilde{G}$-bundle automorphisms (i.e. automorphisms of $\tilde{P}$ covering the 
identity map on the base
 space of this extended bundle or equivalently $T=Maps(M\rightarrow \tilde{G})$). In other words, in this approach one 
wants to make the would-be unified theory 
"look like" a Yang-Mills theory (at least as far as the symmetries are concerned, no mention of the action at 
this point). \textsl{This approach wishes to make use of the crucial advantage of the renormalisability of
 Yang-Mills theories} and thus one hopes that this would-be unified theory (containing gravity) 
will eventually be renormalisable.\\

The first approach works and indeed we have various proposed models and we get information about the underlying 
spacetime geometry dictated by quantum theory. Yet we get no information about the quantization of gravity. 
Concerning the second approach however, it is not even clear whether such a
 group like $\tilde{G}$ exists at all.  We believe that this is one of the
 motivations behind the development of various supergravity or super Yang-Mills theories.  Yet we should be
 careful here: people in supergravity start by "gauging the Poincare group". For simplification we assume
 the Riemannian case and hence the Lorentz group becomes $SO(4)$; this is the structure group of the tangent
 bundle $TM$ of $M$ where by picking a Riemannian metric a reduction of the structure group takes place,
 i.e. we go from $GL(4;{\bf R})$ to $SO(4)$. Yet if one does this, one gets as gauge symmetry group the 
group of bundle automorphisms of the tangent bundle $TM$ of $M$ which cover the identity map on the base 
(the strong tangent bundle automorphisms);
 \emph{clearly, this group is NOT $Diff(M)$, neither does it contain $Diff(M)$}.\\
 
So people add \textsl{fermionic degrees of freedom (Grassmann variables)} (make use of the Coleman-Mandula theorem) 
and now the picture starts becoming messy:
  our understanding is that these grassmann variables are added to the structure group and hence 
one ends up with a super-Lie group.  
\textsl{At the best of our knowledge, there is no proof that even by using this super-Lie group 
as the structure group of a bundle over ordinary spacetime one can get a group of 
strong bundle automorphisms which equals (or contains) $Diff(M)$}. Hence by gauging the Poincare group one 
does not get as group of gauge 
transformations the group of local diffeomorphisms of the spacetime manifold which is the true symmetry group of Einstein's
 general relativity. The advantage however is that this way nonetheless gives indeed renormalisable theories (in fact 
"superrenormalisable theories" i.e. spacetime dimension is \textsl{less} than the upper critical dimension), 
but it also manifests a symmetry between bosons and fermions which does not exist in nature.\\

{\bf Aside:} For completeness we would like to mention the following: in \emph{some} cases 
(namely when the topology of M is such that one can lift a connection of the tangent bundle to a spin 
connection--this is determined by the second Stiefel-Whitney class of M), one can formulate gravity 
via the vierbein and the spin connection). But again, the comments made above between $Diff(M)$ and strong tangent bundle
automorphisms still apply between $Diff(M)$ and the 
group of strong bundle automorphisms of the 
spin bundle (if $dimM=n$, then the spin bundle has structure group the double cover of GL(n,{\bf R}) 
which is the structure group of the tangent bundle $TM$ of $M$).\\
 
 \textsl{We would like to offer 
some ideas of completely different origin on this approach in the next section}.\\

\section{An idea on approximating the group of local diffeomorphisms}

 Let us start with the elementary fact that given a smooth real function
$f:{\bf R}\rightarrow {\bf R}$, we can use Taylor expansion and approximate $f$ by its derivatives (up to infinite order); 
let us now assume that
$f$ is a smooth map from the smooth manifold $M$ onto itself, where $dimM=n$; we pick some local coordinates $\{x^{i}\}$,
 where $i=1,2,...,n$ for $M$; we know that the tangent bundle $TM$ of $M$ has local 
coordinates 
$$\{x^{i},\frac{\partial}{\partial x^{i}}\}$$
 where $dim(TM)=2n$, thus schematically 
$TM$ is like "$M$ plus its first derivative". Similarly, the \textsl{tangent bundle of the tangent bundle} $TTM:=T^{2}M$ will have
 local coordinates containing the $x^{i}$'s, their first and second derivatives. Clearly $dim(T^{2}M)=4n$.
 This is also a bundle over $M$ with structure group $GL(3n;{\bf R})$ since, clearly, the composition of projections
 is again a projection. To approximate a smooth map $f$ from $M$ onto itself then, by immitating the Taylor expansion 
of a real function of 1 real variable, we need to consider 
the \emph{infinite order tangent bundle} $T^{\infty}M$ of $M$ which will be also a bundle over $M$ with
 structure group $GL(\infty ;{\bf R})$. [Note: Strictly speaking the principal bundle is the bundle of linear frames 
of $M$ whose structure Lie group is $GL(n;{\bf R})$ and the tangent bundle $TM$ is its associated vector bundle; hopefully 
there is no misunderstanding caused since we tend not to distinguish between them]. Bundles like $T^{n}M$ appear in
the mathematics literature under the more general title of \emph{jet bundles} (see for instance \cite{djs}).\\  

We know that given in general any algebra (or ring) $A$, we can form the 
group $GL(\infty ;A)$ of invertible $\infty\times\infty$ square matrices with entries from $A$ as follows:
we start with $GL(n;A)$ for some finite $n\in {\bf N^*}$ (i.e. $n$ is a positive integer); there is a 
\textsl{canonical} way to \emph{inject} $GL(n;A)$ into $GL((n+1);A)$: if $C\in GL(n;A)$ is an $n\times n$ invertible 
suqare matrix with entries from the algebra (or ring) $A$, we map it onto the following element in $GL((n+1);A)$:  
$$C\mapsto \sv{C\quad 0}{0\quad 1}$$
Then we take the \emph{inductive limit} of $GL((n+1);A)$ for $n\rightarrow \infty$ which we denote $GL(\infty ;A)$, namely 
one has
$$GL(\infty ;A)=lim _{n\rightarrow \infty}GL((n+1);A)$$

Let us make one remark before proceeding further: clearly the infinite general linear group will be the 
corresponding contribution from the infinite order tangent bundle, so similarly to the Taylor expansion, in order
to approximate the group of local diffeomorphisms we should "add up" the contributions from all orders of the tangent 
bundle; yet the final result will be again, in the limit, $GL(\infty ;A)$.\\
 
There is only one known way to handle this monsterous creature $GL(\infty ;A)$, and this is topology: we can define 
the K-theory groups of $A$ (\textsl{due to Bott perioodicity we have only two of those)} as follows:

$$\pi_{0}[GL(\infty ;A)]:=K^{1}(A)$$
and
$$\pi_{1}[GL(\infty ;A]:=K^{0}(A)$$

[{\bf Aside:}] Using the machinery of the \emph{calculus of functors} (see \cite{goodwill}) in topology, one can indeed 
think of the homotopy groups as being analogous to the derivatives of a smoth function.\\

The main point of this argument is that if one wants to \emph{approximate} smooth maps and hence get a grasp on $Diff(M)$,
 one will
 probably see K-Theory poping up; this seems reasonable since after all K-theory is an $\infty\times\infty$
 generalisation of linear algebra (see for instance \cite{atiyah1}).\\

It is perhaps not clear at this point if one will have to consider $K(M)$, $K({\bf R})$ or its compactification $K(S^{1})$.
 Clearly ${\bf R}$ is 
contractible and noncompact, hence its K-groups are not interesting but we can compactify it to $S^{1}$ (this gives a 
flavour of Kaluza-Klein ideas perhaps) and we know that $K^{0}(S^{1})={\bf Z}$. However the right thing to do, we believe,
 is to consider 
$K(M)$ for the following reason: we know that bundles are locally but not necessarily globally Cartesian products, and 
hence we want 
to consider local and not only global gauge transformations to approximate $Diff(M)$; thus the topology of $M$ should be 
used at some stage. We can be more precise on this point: for convenience we turn from the Lie groups to their 
corresponding Lie algebras: the Lie algebra $b$ of the Lie group $B$ of local gauge transformations $B=Maps(M\rightarrow G)$ 
used above can be expressed as 
$b=g\otimes C(M)$, where $g$ denotes the Lie algebra of the Lie group $G$ and $C(M)$ denotes the (commutative) algebra 
of functions on the manifold $M$, namely we consider 
matrices in $g$ with entries from $C(M)$; that amounts to, in the above discussion, taking $A=C(M)$, 
(namely we replace ${\bf R}$ with 
$C(M)$ and $GL(n;{\bf R})$ with $GL(n;C(M))$ since we need the general linear group  as the strucutre group of the 
tangent bundle and its powers), 
hence if we take the 
inductive limit and then take its fundamental group we shall end up 
with the K-groups of the algebra $C(M)$; but Serre-Swan theorem tells us that this is equal to the topological K-theory 
of the manifold $M$ which is what we considered.\\

The bottom line of this argument is that following 
the second way (namely try to make the unified theory look like a Yang-Mills theory for which we have a good
understanding of quantization and renormalisation), the sought after extended group should be 
$\tilde{G}=K(M)\times G$,  where $G$ is the "honest" Lie group $G=SU(3)\times U(2)\times SO(4)$ 
for the 
strong, electroweak and "linear gravity" interactions respectively; yet the "total" group $\tilde{G}$ should contain the 
semi-direct product 
 with an additional \emph{discrete} group, the K-theory group $K(M)$ of the spacetime manifold $M$ which would take 
care of the \emph{"nonlinear" part of the local diffeomorphisms}.\\
 
It is perhaps more \textsl{convenient} to take the \textsl{crossed product
 noncommutative algebra} $D=K^{0}(M)\rtimes C(G)$ where $C(G)$ denotes the commutative algebra of functions on the Lie 
group $G$ seen as a manifold (in which case we are not considering the Lie group structure on $G$). Since $D$ is a 
noncommutative algebra, one can very easily turn that 
into a Lie algebra by taking the commutator of two elements as the Lie bracket. Hence we have a 
\textsl{good candidate at least 
for the} \emph{adjoint bundle} of the sought for principal $\tilde{G}$-bundle. Thus one can define connection 1-forms 
(gauge potentials) and curvature 2-forms (gauge fields) since these are Lie algebra valued. However this will not be a 
Lie algebra coming necessarily  from a Lie group, at least not in a straitforward way. Perhaps there is an underlying 
quantum Lie group yet to be discovered. The principal bundle itself, apart from providing the  
finite gauge transformations (and not only the infinitesimal ones as the adjoint bundle does), is important for an 
additonal reason: the holonomy of the connection on the $\tilde{G}$-bundle over spacetime, (which physically corresponds 
to the Dirac phase factor of the potential which is the true 
quantum observable from the Aharonov-Bohm effect),
 is an element of the structure group of the bundle $\tilde{G}$.\\

 \textsl{Clearly we end up again with a} 
\emph{noncommutative} space, since $K(-)$ is a discrete group crossed product with the commutative algebra of 
functions on an honest Lie group $C(G)$. Hence 
it appears that whichever of the two obvious approaches one follows for a unified theory of all interactions, 
noncommutative 
geometry enters the scene either as a noncommutative spacetime or as a noncommutative algebra of some underlying 
structure group 
(perhaps a quantum Lie group).\\
 
It is fairly clear we belive from the above discussion that this approach involving K-theory gives a better 
approximation of $Diff(M)$ than supersymmetry, at least topologically.\\

In order to define the crossed product algebra between the K-group and say $C(G)$, we 
need an action of the (discrete) group $K^{0}(M)$ onto $G$. This can be defined, for example, by using the 
\emph{holonomy} of a connection,
 in a similar fashion used in \cite{nlsm}, as follows:
let $P$ be the total space of a principal $G$-bundle over $M$ and we assume that $M$ is not simply connected; 
we know that gauge equivalent classes of flat connections are in one to one correspondence with 
conjugacy classes of irreducible representations of $\pi _{1}(M)$ onto $G$. Let $A$ be a representative
of such a class and let us assume that $A$ has holonomy. The holonomy of $A$ defines a map $h: \pi _{1}(M)\rightarrow G$.
Let $H$ denote the image of $\pi _{1}(M)$ into $G$ under $h$, namely $H=h(\pi _{1}(M))$ which is a subgroup of $G$. Thus
we have an action of $\pi _{1}(M)$ onto $G$ which is defined by the usual multiplication in $G$ restricted to $H$. 
But we know that $K^{0}(M)=\pi _{1}(GL(\infty ;C(M)))$, where $C(M)$ denotes the commutative ring of functions on $M$.
Thus we have an induced "holonomy map". One can relax the flatness
condition on the connection $A$; in this case, provided that holonomy exists, one still gets a representation of 
$\pi _{1}(M)$ onto $G$ but this representation is more complicated and it may not be irreducible. The physical 
picture is that since we are dealing with gravity where other gauge fields are present, picking a gauge class of 
connections (potentials), corresponds to a choice of a, say, $\theta $ vacuum.\\ 

Let us close this section with some remarks:\\

{\bf 1.} There is a point which is still unclear: why should we take only the fundamental group of $GL(\infty ;A)$ and not 
all its 
homotopy groups? We know that due to Bott periodicity there are only two K-groups; however one can also use the higher 
homotopy groups provided one applies Quillen's famous \emph{plus} construction which will give nontrivial higher K-groups.
But the role of these higher K-groups is unclear even in the mathematics literature. Another option would be to take 
the group ring of $\pi _1(M)$ and apply Quillen's ideas to it; this will lead us to the Waldhausen K-theory 
(see \cite{atiyah1}) but it is not easy to relate the Waldhausen K-groups to physics which is what we are trying to do
 in this article.\\

{\bf 2.} There are some more versions of "supersymmetric" theories where the Grassmann variables are added to the spacetime
 manifold as extra degrees of freedom. The philosophy of this approach looks more like the attempt to make the unified theory 
look like a gravity theory on a "graded commutative" space. Supersymmetry is quite popular in the physics community and since 
at least until now, there is no experimental evidence for its existence, people assume that it must be spontaneously broken.\\

{\bf 3.} There has been around in the literature for 15 years or so the notion of \emph{quantum bundles} 
(through the work of Majid etc, see for example \cite{majid}); one can, in a sense, say that what we propose here is 
some sort of a quantum bundle 
structure over spacetime where the structure group is not a quantum Lie group (which is what is used in the definition 
of quantumm bundles; a quantum Lie group comes from deformations of classical Lie groups) but it is a 
noncommutative space (defined via its noncommutative algebra of coordinate functions).\\

{\bf 4.} Since a Riemannian metric reduces the structure group of the tangent bundle from $GL$ to $SO$, perhaps one 
should take $KO$-groups instead of $K$-groups.\\

\section{Discussion}

Our motivation for this article came from an attempt, eventually, to see if one can say something useful about the problems 
of renormalisation of quantum gravity, especially under the light of the recent work of Connes and Kreimer (see 
\cite{ck1}, \cite{k} and \cite{bk}). We are not in a position to do that yet but we shall try to make some 
comments to motivate further research.\\

Renormalisation has its origin in the Kopenhagen interpretation of quantum mechanics as a probabilistic theory; 
it is absolutely crucial in quantum field theory in order to relate the \emph{bare} quantities of the theory (those are 
the parameters appearing in the action), with \emph{physical} quantities (those measured in an accelerator during an 
experiment). The later depend on the energy scale in which each experiment is conducted [thus there is an action of the 
multiplicative group ${\bf R}_{+}$ on the space of all physical parameters; this space is often a manifold and the group 
action is not always free, it may have fixed points]; 
moreover many of them appear to be infinite; in many cases this is inherited by the classical theory (e.g. like the 
charge of the electron in QED: it is infinite classically if we consider the electron as a point particle with the 
inverse square law and it remains infinite in QED prior to renormalisation); in fact renormalisation at first it was 
considered as an ingenious (but rather \textsl{ad hoc}) way to extract finite answers by using a 
\emph{finite} sypply of countertems which cancel the infinities (divergent integrals) appearing in the perturbation 
series expansion of the action using Feynman graphs, and at the same time all quantities become independent 
from the energy scale at the end of the calculation. In all this, gauge symmetry (via Ward-Takahasi and 
Slavnov-Taylor equations for the abelian and non-abelian cases respectively) 
plays a key role. Recent developments related to the renormalisation group flow however have pointed out that 
renormalisation is something really deep in physics and it represents a lot more than a number of clever techniques 
(eg the qualitative explanation of the finiteness of electric charge of the electron by "screening", uncertainty 
principle and polarisation of the vacuum).\\

It has been observed that if the Lagrangian contains combinations of field operators of excessively high dimension in 
energy units compared with the spacetime dimension, the counterterms required to cancel all divergenies proliferate to 
\emph{infinite} number, and, at first glance, the theory would seem to gain an infinite number of free parameters and 
therefore it loses all predictive power. This does not happen for electroweak and strong interactions (although in general 
this is not the end of the story since there may be anomalies), but it seems to happen 
in gravity if one applies perturbation theory in it. In such a case, there are very few escapes if one is lucky: 
gauge symmetry (that is why we care so 
much about the gauge group of symmetries), the so-called \emph{Fisher-Wilson fixed points} in the renormalisation 
group flow or nonperturbative effects.\\

Before elaborating more on these Fisher-Wilson fixed points, let us try to recall the current level of affairs about 
perturbative quantum gravity (see \cite{vp} and \cite{hw}) (but at the same time
 we shall make a more general study): we focus on Newton's
constant $G$; in general, it depends on the energy scale; using physical units as usual we translate length into the inverse
 of momentum. This allows one to think $G$ as a function of momentum $p$, namely $G=G(p)$ (higher momentum means shorter 
distances). The  equation describing how $G$ depends on $p$ is the \textsl{Callan-Symanzik} equation:
$$\frac{dG}{d(lnp)}=\beta (G)$$
which involves the $\beta $-function of our theory. Typically $\beta (G)=(n-d)G+aG^{2}+bG^3+...$ where $n$ is spacetime 
dimension and $d$ is the \emph{upper critical dimension} (this is the spacetime dimension in which the coupling constant 
is dimensionless). Let's take the linear term: 
$$\frac{dG}{d(lnp)}=(n-d)G$$ 
which means that 
$G$ is proportional to $p^{n-d}$. One can study 3 cases:\\
{\bf (i)} if $n<d$, then the Newtonian constant gets \textsl{smaller} at higher momenta which means that in higher 
energy scales there is practically no interaction and the theory is essentially \emph{free} (superrenormalisable theory);\\
{\bf (ii)} if $n>d$, then Newton's constant gets larger at higher energies, so gravity becomes very strong 
(in fact infinitely 
strong) as momentum increases. In this case the theory is (superficially perturbatively) nonrenormalisable; 
things are very difficult but there may 
be some hope in some cases;\\
{\bf (iii)} if \emph{n=d}, we say the theory is renormalisable (like QED) but we have to calculate the next term in the 
$\beta $-function.\\

We are interested in quantum gravity in 4 dimensions, namely $n=4$. So we have to figure out what $d$ is. Well, a not 
very hard argument about dimension and units says that $d=2$ (see for example \cite{vp}, whereas the upper critical 
dimension for Yang-Mills theories is 4). 
So gravity in 4 dimensions is (superficially at 
least) nonrenormalisable. Yet there is a subtlety here: we should take care of the higher order terms of the 
$\beta$-function.\\

Let's focus on the second term; this becomes dominant if $n=d$, ie for renormalisable theories (like QED):
$$\frac{dG}{d(lnp)}=aG^{2}.$$ 
One can solve this easily and get 
$$G=\frac{c}{1-aclnp}$$ 
where $c$ is a positive constant. If 
$a<0$ the coupling constant slowly decreases with increasing momentum, in this case we say that our theory is 
\emph{asymptotically free} (like QCD).\\

If however $a>0$, then $G$ increases as the energy scale goes up, in fact it bocemes infinite at sufficiently high energies,
this is called a \textsl{Landau pole} (this happens in QED when we do not include the weak force). Among renormalisable 
theories the ones with $a<0$ are considered "good" and the ones with $a>0$ are considered "bad" (like QED).\\

Let's come to gravity now: perturbative quantum gravity in dim 2 is not only renormalisable, in fact it is 
asymptotically free, since for gravity $a<0$. If we ignore higher order terms, this implies something very interesting 
about gravity in 4 dimensions: if we see only the first 2 terms, then $G$ increases as the momentum increases (as an
honest nonrenormalisable theory would do) but when $G$ gets big enough, the second term matters more (remember it has 
a negative coefficient in front of it) and thus after a while the growth of $G$ starts slowing!\\
 
There is \emph{strong numerical evidence} (see \cite{hw}) that in fact
$$lim _{p\rightarrow \infty}\frac{dG}{d(lnp)}\rightarrow 0$$
This is called an \textsl{"ultraviolet stable fixed point"}. Mathematically, it attracts nearby points as we flow in the 
direction of higher momenta. This particular kind of ultraviolet stable fixed point-coming from an asymptotically free 
theory in dimensions above its upper critical dimension-is called a \emph{"Fisher-Wilson" fixed point} (see \cite{vp}).\\

In general $\beta $-function computations are hard. The big question then is the following: can we use the new 
Connes-Kreimer approach to $\beta$-function using the
Hopf algebra of Feynman graphs (or equivalently the Hopf algebra of decorated rooted trees, see 
\cite{ck1}, \cite{k} and \cite{bk}) 
in order to prove that the $\beta $-function of perturbative quantum gravity has a Fisher-Wilson fixed point?\\

\section{Appendix}

Let us for convenience, describe briefly the Connes-Kreimer approach to renormalisation; \textsl{given a specific 
quantum field theory}, namely its action and its symmetries, (although in \cite{k1} the author managed to describe a 
generic quantum Yang-Mills theory purely combimatorial without starting from a gauge invariant Yang-Mills type of action), 
we need the following data 
(in momentum space, a similar description exists in coordinate space):
$(H,V,R, \phi )$ where $H$ is the Hopf algbra of Feynman graphs (equivalntly one can consider the Hopf algebra of 
decorated rooted trees), $V$ is the regularisation algebra (for dimensional regularisation 
$V={\bf C}[\epsilon ^{-1}, \epsilon ]$, ie $V$ is the algebra of Laurent series with finite pole part), 
$\phi :H\rightarrow V$ is a (umital) algebra homomorphism, the "regularisation map"which can be seen as corresponding 
to a choice of a boundary condition for the Dyson-Schwinger equation (regularised Feynman rules, see \cite{bk}) and 
$R: V\rightarrow V$ is the "renormalisation scheme": it satisfies the Rota-Baxter equation (this equation provides the 
link between renormalisation and the
Birkhoff decomposition from which one can get the Riemann-Hilbert correspondence), it has $R(1)=1$ and it 
preserves the UV divergent structure (ie the pole part--e.g. in the minimal subtraction scheme, $R$ is the projector onto 
the proper part). The Hopf algebra 
$H$ has a coproduct $\Delta $ (which disentagles trees and divergencies to subtrees and subdivergencies) and an antipode 
$S$. 
One then defines the twisted antipode $S^{\phi}_{R}: H\rightarrow V$ which provides the relevant counterterm and 
the convolution $S^{\phi}_{R}*\phi =\phi _{R}$ which solves the Bogoliubov recursion. The Hopf algebra structure can be 
determined by the perturbative expansion of the action into Feynamn graphs; one can have two models for $H$: either 
the graded, free, commutative algebra generated by trees (by a tree we mean a connected, contractible, compact graph) 
with the 
weight (which is the number of vertices) grading, or, given a set $S$, the graded, free, commutative algebra generated by 
$S$-decorated trees (we decorate only the vertices and not the edges of the graphs). All Hopf algebras appearing in various
quantum field theories are Hopf subalgebras of the two models above. Note that we can base the Hopf algebra on 1PI 
(1 particle irreducible) Feynman 
graphs instead of trees, these are equivalent descriptions. We are interested in the Hochschild cohomology of the 
Hopf algebra of trees $H$, in fact since Hochschild cohomology groups 
vanish in dimension greater than 2, we are only interested in the first Hochschild cohomology groups. This is crucial, 
since in the 
Connes-Kreimer framework, the requirement of the locality of the counterterms in renormalisation is interpreted as 
linear functionals on $H$ being \emph{$b$-closed}, where $b$ is the Hochschild differential (see \cite{k1} and \cite{bk}).\\

To the Hopf algebra of graphs $H$ one can associate a Lie algebra and a Lie group, let's denote it $G$, using the
 Milnor-Moore theorem. Roughly $G$ comes from the group of characters of $H$. The antipode
 map $S$ in the Hopf algebra delivers the same terms as those needed for the subtraction procedure in 
renormalisation. One can understand $S$ by using the Riemann-Hilbert correspondence.\\

 A well-known instance of the Riemann-Hilbert correspondence (which in general gives equivalences between
 geometric problems associated with differential systems with singularities and representation theoretic data)
 is the 1:1 correspondence between (gauge equivalence classes of) \emph{flat} connections on a vector
 bundle and (conjugacy classes of) representations of the fundamental group of the base manifold onto the
 strucure Lie group, the correspondence given by the \emph{holonomy} of the flat connection. \\ 

Essentially Connes-Kreimer used a variation of the above well-known example of the Riemann-Hilbert correspondence: 
the base is an infinitesimal punctured disc (which is a non-simply connected space)
 $\Delta ^{*}$ of ${\bf CP}^{1}$
 around the point $D=4$ (the complex surface comes from the complexification of dimension which from ${\bf N}$ 
takes values in ${\bf R}$ according to the rules of dimensional regularisation where dimension becomes real
 and then we complexify), the 
strucure group is ${\bf G_{m}}$ which is just the complex numbers (as multiplicative group)
 and the total space is denoted $B$. The fibre represents \emph{rescaling} and then we study equisingular 
(representing independence of choice of unit of mass) $G$-valued   flat connections (the Lie group $G$ is 
the one coming from group of characters of the Hopf algebra $H$ of Feynman graphs due to Kreimer). From the 
representation theory side we study  representations $U^{*}\rightarrow G^{*}$ where $U$ is the universal group of all 
physical theories and $G^{*}=G\rtimes _{\theta} {\bf R}$ (see \cite{cm}).\\

 Now let us try to duplicate the Connes-Kreimer framework in the case of gravity, namely we would like to build the 
Riemann-Hilbert correspondence for quantum gravity. Clearly $U^{*}$ is a universal group for all physical theories, 
hence there is no change here. Similarly the punctured disc $\Delta ^{*}$ should remain the same for a possible
 dimensional regularisation of quantum gravity in dim 4. The structure group ${\bf G}_{m}$ which representrs rescalling
 should not be changed either. Thus the only ingredient which changes for the case of quantum gravity is the 
Connes-Kreimer group $G$ since gravity has different action and different group of gauge transformations.\\

We would like to finish this section with the following remark: if we manage to prove that
 the $\beta $-function of Newton's constant for quantum gravity has a Fisher-Wilson fixed point, (using the Connes-Kreimer 
framework or other), that would definitely be a major breakthrough. Yet this would not be the end of the story
since a proper quantum theory of gravity should explain the cosmological constant puzzles as well.
Moreover, quantum gravity is a theory which has also, crucially, \emph{nonlocal features} (e.g. holography principle); so 
relying solely on the 
Hochschild cohomology of the Hopf algebra is too restrictive; somehow, following the line of argument in 
\cite{konts},  \cite{z} and
the Connes-Kreimer approach to renormalisation,
this may be related to operads and the action of the Grothendieck-Teichmuller group on the Hochschild complex.
But there is another feature that a possible quantum gravity theory must have: it must be \emph{asymmetric in time} in 
order
to explain the difference in entropy between the big-bang and the big-crunch (see \cite{rp}). Unfortunately both 
general relativity 
and quantum mechanics are theories which are symmetric in time. That, perhaps, would require a more radical approach 
to quantum gravity than
simply trying to immitate perturbative quantum field theory ideas.\\

\emph{Acknowledgement:} The author wishes to thank A. Connes, D. Kreimer, D.G. Quillen, R. Penrose and G. Segal for 
useful discussions.\\


\begin{thebibliography}{20}





\bibitem{atiyah1} M.F. Atiyah: \emph{"K-Theory"}, Benjamin 1967.\\

D.G. Quillen: \emph{"Higher Algebraic K-Theory"}, Graduate Lecture Course, Oxford, Michaelmas 1994 (personal notes).\\

\bibitem{atiyah2}M.F. Atiyah: \emph{Collected Works (Vol 5: Gauge Theory)}, Oxford University Press (1990).\\ 

M.F. Atiyah: \emph{"The Geometry and Physics of Knots"}, Accademia Nazionale dei Lincei, Cambridge University Press, 1990.\\

\bibitem{connes} A. Connes: \emph{``Noncommutative Geometry''}, Academic Press, (1994).\\

\bibitem{djs} D. J. Saunders: \emph{The Geometry of Jet Bundles}, London Mathematical Society Lecture Note Series 142,
Cambridge University Press, (1989).\\

\bibitem{majid} S. Majid: \emph{"Riemannian geometry of quantum groups and finite groups with nonuniversal differentials"},
 Commun. Math. Phys. 225 (2002) 131-170.\\

\bibitem{ck1} A. Connes and D. Kreimer: \emph{"Hopf Algebras, Renormalisation and Noncommutative Geometry"}, Commun. Math. 
Phys. 199, (1998), 203-242.\\

A. Connes, D. Kreimer: \emph{``Renormalization in quantum field theoy and the Riemann-Hilbert problem.
I. The Hopf algebra structure of graphs and the main theorem"}, Commun. Math. Phys. 210.1 (2000), 249-273.\\

A. Connes, D. Kreimer: \emph{"Renormalization in quantum field theoy and the Riemann-Hilbert problem.
II. The $\beta$-function, diffeomorphisms and the renormalisation group"}, Commun. Math. Phys. 216.1 (2001), 215-241.\\

\bibitem{connes1}A. Connes, D. Kreimer: \emph{"Lessons from Quantum Field Theory"}, hep-th/9904044.\\

A. Connes and D. Kreimer: \emph{"Renormalisation in 	Quantum Field Theory and the Riemann-Hilbert problem"}, 
hep-th/9909126.\\

\bibitem{cm}A. Connes and M. Marcolli: \emph{"Quantum fields and motives"},  J.Geom.Phys. 56 (2006) 55-85.\\

\bibitem{k} D. Kreimer: \emph{"On the Hopf algebra structure of perturbative Quantum Field Theory"}, 
Adv. Theor. Math. Phys. 2 no 2 (1998), 303-334.\\ 

D. Kreimer: \emph{"Chen's iterated integral represents the operator product expansion"}, Adv. Theor. Math. Phys. 3.3
(2000).\\

\bibitem{k1}D. Kreimer: \emph{"Anatomy of a Gauge Theory"}, hep-th/0509135.\\

\bibitem{bk} C. Bergbauer and D. Kreimer: \emph{"Hopf Algebras in Renormalization Theory: Locality and Dyson-Schwinger 
Equations from Hochschild Cohomology"}, hep-th/0506190.\\

\bibitem{konts} M. Kontsevich: \emph{"Operads and Motives in Deformation Quantization"}, 
Lett. Math. Phys. 48.1, (1999), 303-334.\\

\bibitem{z} I.P. Zois: \emph{"Operads and Quantum Gravity"}, Repts on Math. Phys. 55.3, (2005), 307-323.\\

\bibitem{nlsm} I. P. Zois: \emph{"A new invariant for $\sigma $-models"}, Commun. Math. Phys. 209.3, (2000), 757-783.\\ 

\bibitem{iz} C. Itzykson, J-B. Zuber: \emph{"Quantum Field Theory"}, McGraw-Hill, 1980.\\ 

J. Collins: \emph{"Renormalisation"}, Cambridge Monographs in Mathematical Physics, Cambridge University Press, 1984.\\

S. Pocorski: \emph{"Gauge Field Theories"}, Cambridge University Press, 1987.\\

P. Ramond: \emph{"Field Theory: A Modern Primer"}, Addison-Wesley, 1990.\\

M. E. Peskin and D. V. Schroeder: \emph{"An Introduction to Quantum Field Theory"}, Addison-Wesley, 1995.\\ 

J. Zinn-Justin: \emph{"Quantum Field Theory and Critical Phenomena"}, Oxford University Press 1993.\\

\bibitem{vp}S. Weinberg: \emph{"Ultraviolet divergencies in quantum theories of gravitation"}, in \textsl{General 
relativity, an Einstein Centenary survey}, S. W. Hawking, W. Israel (eds), Cambridge University Press (1979).\\

M. Fisher and K. Wilson: \emph{"Critical exponents in 3.99 dimensions"}, Phys. Rev. Lett. 28 (1972), 240.\\

D. Gross and F. Wilczek: \emph{"Ultraviolet behavior of nonabelian gauge theories"}, Phys. Rev. Lett. 30 (1973), 1343.\\

V. Periwal: \emph{"Cosmological and astrophysical tests of quantum gravity"}, astro-ph/9906253.\\

\bibitem{hw}H. W. Hamber and R. M. Williams: \emph{"Newtonian potential in quantum Regge gravity"}, Nucl. Phys. B435 
(1995), 361-397.\\

O. Lauscher and M. Reuter: \emph{"Is Quantum Einstein Gravity Nonperturbatively Renormalisable?"}, Class. Quant. Grav. 19 
(2002), 483-492.\\

O. Lauscher and M. Reuter: \emph{"Towards Nonperturbative Renormalisability of Quantum Einstein Gravity"}, Int. J. Mod.
 Phys. A17 (2002), 993-1002.\\

O. Lauscher and M. Reuter: \emph{"Fractal Spacetime Structure in Asymptotically Safe Gravity"}, hep-th/0508202.


\bibitem{rp}R. Penrose: \emph{"The Road to Reality"}, Oxford University Press 2003 and \textsl{private communication}.\\

\bibitem{goodwill} T. Goodwillie: \emph{"Calculus of Functors"}, LMS Summer School on the Calculus of Functors and its 
Applications, University of Aberdeen, UK, (2000), (personal notes).\\






\end{thebibliography}
\end{document}